\documentclass[fleqn,usenatbib]{mnras}
\usepackage[T1]{fontenc}
\usepackage{ae,aecompl}

\usepackage{graphicx}	
\usepackage{amsmath}	
\usepackage{amssymb}	
\usepackage{newtxtext,newtxmath}

\title[AS2UDS: Clustering of SMGs]{An ALMA survey of the SCUBA-2 Cosmology Legacy Survey UKIDSS/UDS field: Halo Masses for Submillimetre Galaxies}

\author[Stach et al.]{S.\,M.\ Stach,$^{1}$ $\!$\thanks{E-mail: stuart.m.stach@durham.ac.uk}
I.\ Smail,$^{\! 1}$
A.\,Amvrosiadis,$^{1}$
A.\,M.\ Swinbank,$^{\! 1}$
U.\ Dudzevi\v{c}i\={u}t\.{e},$^{\! 1}$
\newauthor
J.\,E.\ Geach,$^{\! 2}$
O.\ Almaini,$^{\! 3}$
J.\,E.\ Birkin,$^{\! 1}$
Chian-Chou Chen,$^{\! 4}$
C.\,J.\ Conselice,$^{\! 3}$
\newauthor
E.\,A.\ Cooke,$^{\! 1,5}$
K.\,E.\,K.\ Coppin,$^{\! 2}$
J.\,S.\ Dunlop,$^{\! 6}$
D.\ Farrah,$^{\! 7,8}$
S.\ Ikarashi,$^{\! 1}$
\newauthor
R.\,J.\ Ivison,$^{\! 9}$
J.\,L.\ Wardlow$^{\! 10}$
\\
$^{1}$Centre for Extragalactic Astronomy, Department of Physics, Durham University, Durham, DH1 3LE, UK\\
$^{2}$Centre for Astrophysics Research, School of Physics, Astronomy and Mathematics, University of Hertfordshire, Hatfield AL10 9AB, UK\\
$^{3}$School of Physics and Astronomy, University of Nottingham, University Park, Nottingham, NG7 2RD, UK\\
$^{4}$Academia Sinica Institute of Astronomy and Astrophysics, No.\ 1, Sec.\ 4, Roosevelt Rd., Taipei 10617, Taiwan\\
$^{5}$National Physical Laboratory, Hampton Road, Teddington, Middlesex, TW11 0LW, UK\\
$^{6}$Institute for Astronomy, University of Edinburgh, Royal Observatory, Blackford Hill, Edinburgh EH9 3HJ, UK\\
$^{7}$Department of Physics and Astronomy, University of Hawaii, 2505 Correa Road, Honolulu, HI 96822, USA\\
$^{8}$Institute for Astronomy, 2680 Woodlawn Drive, University of Hawaii, Honolulu, HI 96822, USA\\
$^{9}$European Southern Observatory, Karl Schwarzschild Strasse 2, Garching, Germany\\
$^{10}$Department of Physics, Lancaster University, Lancaster, LA1 4YB, UK\\
}

\date{Accepted ---. Received ---; in original form ---}

\pubyear{2020}

\begin{document}
\label{firstpage}
\pagerange{\pageref{firstpage}--\pageref{lastpage}}
\maketitle
\begin{abstract}
We present an analysis of the spatial clustering of a  large sample of high-resolution, interferometically identified, submillimetre galaxies (SMGs). We measure the projected cross-correlation function of $\sim$\,350 SMGs in the UKIDSS Ultra Deep-Survey Field across a redshift range of $z$\,$=$\,1.5--3 utilising a method that incorporates the uncertainties in the redshift measurements for both the SMGs and cross-correlated galaxies through sampling their full probability distribution functions. By measuring the absolute linear bias of the SMGs we derive halo masses of $\log_{10}(M_{\rm halo}[{h^{-1}\,\rm M_{\odot}}])$\,$\sim$\,12.8 with no evidence of  evolution in the halo masses with redshift, contrary to some previous work. From considering models of halo mass growth rates we predict that the SMGs will reside in haloes of mass $\log_{10}(M_{\rm halo}[{h^{-1}\,\rm M_{\odot}}])$\,$\sim$\,13.2 at $z$\,$=$\,0, consistent with the expectation that the majority of $z$\,$=$\,1.5--3 SMGs will evolve into present-day  spheroidal galaxies. Finally, comparing to  models of stellar-to-halo mass ratios, we show that SMGs may correspond to systems that are maximally efficient at converting their gas reservoirs into stars.  We compare them to a simple model for gas cooling in halos that suggests that the unique properties of the SMG population, including their high levels of star-formation and their redshift distribution, are a result of the SMGs being  the most massive galaxies that are still able to accrete cool gas from their surrounding intragalactic medium.
\end{abstract}


%
%
%
\section{Introduction}

Submillimetre galaxies (SMGs) are a population of high-redshift dusty galaxies \citep[typically $z$\,$\sim$\,2--3:][]{chapman2005redshift,chapin2009aztec,amblard2010herschel,simpson2014alma,danielson2017alma,dudzevivciute2020alma}, with far-infrared luminosities ($L_{\rm IR}$) $\sim$\,10$^{12-13}$\,L$_{\odot}$ \citep[see:][for review]{casey2014dusty}. It is believed that the majority of this far-infrared emission corresponds to  dust-reprocessed radiation from recent star formation, with the luminosity of this emission implying high dust masses ($\gtrsim$\,10$^{8}$\,M$_{\odot}$), and high star-formation rates ($>$\,100\,M$_{\odot}$\,yr$^{-1}$), and thus SMGs are some of the most massive and rapidly star-forming galaxies in the Universe. In addition, their 
selection at submillimetre wavelengths ($\sim$\,850--1250\,$\mu$m) corresponds to the Rayleigh-Jeans tail of the galaxy spectral energy distribution (SED) in high-redshift galaxies, which results in a strongly negative $k$-correction \citep[figure 4:][]{blain2002submillimeter}. As a result, at a fixed observed wavelength in the submillimetre, as the redshift of an SMG is increased the SED is sampled along the rising Rayleigh-Jeans tail and this increasing brightness approximately cancels out the luminosity dimming from the increasing distance out to $z$\,$\sim$\,7. Therefore, 
submillimetre surveys for SMGs provide effectively volume-limited
probes of strongly star-forming galaxies with high dust mass and by implication high gas mass, in the high-redshift Universe.

Given the estimated gas masses and star-formation rates of SMGS \citep[e.g.][]{bothwell2013survey,dudzevivciute2020alma}, their extreme star formation rates can only be a relatively short lived  \citep[$\sim$\,200\,Myr e.g.][]{birkin2020alma} and much work has been undertaken to understand where this infrared-bright phase fits into a larger evolutionary pathway for SMGs. One suggestion is the \cite{sanders1988ultraluminous} scenario for the local Universe analogues: ultra-luminous infrared galaxies (ULIRGs, with $L_{\rm IR}$\,$\geq$\,10$^{12}$\,L$_\odot$), which places 
the strongly star-forming ULIRGs as an intermediate phase following a galaxy merger and preceding a resultant quasar phase, with the present-day descendant being a massive passive spheroidal galaxy. There is circumstantial evidence for this link from the redshift distributions of SMGs and quasars which peak at similar redshifts \citep{chapman2005redshift, wardlow2011laboca, assef2011mid, dudzevivciute2020alma}. In addition a number of other observational tests are claimed to support this evolutionary link \citep[e.g.][]{swinbank2006link, tacconi2008submillimeter,hainline2011stellar, simpson2014alma, simpson2017scuba, hodge2016kiloparsec, stach2019alma, dudzevivciute2020alma}.
However, these tests are uncertain as they  rely on measurements and models that are poorly constrained e.g.\ stellar masses and star-formation histories \citep[e.g.][]{hainline2011stellar, michalowski2012stellar}.

Another method for contextualising the evolution of a galaxy population is through measurements of their spatial clustering, which is linked to the masses of their dark matter haloes \citep{peebles1980large}. With inferred dark matter halo masses, the present-day halo masses for a galaxy population can be estimated based on the dark matter mass assembly histories from N-body simulations \citep{fakhouri2010merger}.  Using these, comparisons can then be made with clustering measurements of the proposed evolutionary descendants in the local Universe. This spatial clustering method has been applied to SMGs \citep[e.g.][]{blain2004clustering,weiss2009large,cooray2010hermes,lindner2011deep,hickox2012laboca,chen2016faint,wilkinson2016scuba,amvrosiadis2019herschel} and where clustering signals could be detected there is general agreement that SMGs reside in massive dark matter haloes of mass $M_{\rm halo}$\,$\sim$\,10$^{12-13}$\,M$_{\odot}$. This halo mass is broadly in agreement with those expected for an evolutionary track connecting QSOs \citep{croom20052df,myers2006first,hickox2011clustering} and local massive spheroidal galaxies \citep{quadri2007clustering,zehavi2011galaxy}, supporting a \cite{sanders1988ultraluminous}-like evolutionary model.

The major difficulties with measuring the clustering signal for SMGs are the relative low number densities of SMGs, resulting in small sample sizes, and the reliance on uncertain identification and similarly uncertain photometric redshifts, due to the challenges and expense of obtaining spectroscopic redshifts for SMGs.  \cite{hickox2012laboca}  attempted to minimise these issues by cross-correlating a small sample of probable SMGs (some with spectroscopic redshifts) in the Extended \textit{Chandra} Deep Field-South, with a larger galaxy  sample in the same field from the \textit{Spitzer} InfraRed Array Camera (IRAC). In addition they adopted the \cite{myers2009incorporating} method for incorporating the information in the full probability distribution function (PDF) for the photometric redshift estimations for the IRAC galaxies to improve the resulting clustering signal. With this method they derived an auto-correlation length for $\sim$\,50 SMGs in the redshift range $z$\,$=$\,1.5--3 of $r_{0}$\,$=$\,7.7$^{+1.8}_{-2.3}$\,$h^{-1}$\,Mpc which corresponded to a dark matter halo mass of $\log_{10}(M_{\rm halo}[{h^{-1}\,\rm M_{\odot}}])$\,$=$\,12.8$^{+0.3}_{-0.5}$.

More recently, larger submillimetre samples have become available, as degree-scale extra-galactic fields have been mapped at submillimetre wavelengths with single-dish facilities, increasing the precision of clustering measurements \citep{chen2016faint,wilkinson2016scuba,amvrosiadis2019herschel,an2019multi,lim2020scuba}. These larger surveys allow the samples to be split by redshift  to measure the evolution in clustering strength as a function of redshift, however currently there are disagreements about the trends found. \cite{wilkinson2016scuba} found redshift evolution such that SMG activity occurs in more massive dark matter halo masses ($M_{\rm halo}$\,$\sim$\,10$^{13}$\,M$_{\odot}$) at higher redshifts ($z$\,$>$\,2) and in lower mass haloes ($M_{\rm halo}$\,$\sim$\,10$^{11}$\,M$_{\odot}$) at $z$\,$<$\,2. Contrary to this, the observed clustering measurements from \cite{chen2016scuba}, \cite{amvrosiadis2019herschel}, and \cite{an2019multi} \citep[as well as results from  semi-analytical models of][]{cowley2016clustering} suggest SMGs inhabit haloes of $M_{\rm halo}$\,$\sim$\,10$^{12}$\,M$_{\odot}$ at all redshifts. 

Finding the source of this disagreement is complicated by the differing methods used to identify SMGs from the low-resolution single-dish maps, which are known to suffer from source blending \citep[e.g.][]{karim2013alma,stach2018alma}. All of these studies rely on probabilistic radio, mid-infrared and colour-selection for identifications that are known to be incomplete and contaminated \citep{hodge2013alma}. Any mis-identification of the SMGs can have dramatic effects on the resulting clustering measurements and could be responsible for the conflicting claims about the halo mass evolution. For example, with the availability of robust identifications for samples of SMGs using the Atacama Large Millimetre/submillimetre Array (ALMA), \cite{garcia2020clustering} has suggested that single-dish clustering studies could be overestimating the SMG halo masses by as much as 3.8$^{+3.8}_{-2.6}$ times their true value. Therefore, to obtain robust results such an analysis needs to be based on a large sample of SMGs across a contiguous field that are accurately identified through submillimetre interferometry at sub-arcsecond resolution to yield a precise and accurate measurement of SMG halo masses.

We have recently completed an ALMA follow-up survey of the $\sim$\,700 submillimetre sources in the 850-$\mu$m map of the UKIDSS Ultra Deep Survey (UDS) field obtained by the SCUBA-2 Cosmology Legacy Survey \citep[S2CLS,][]{geach2017scuba}. The S2CLS UDS map reached a median sensitivity of $\sigma_{850}$\,$=$\,0.9\,mJy over an area of 0.96\,deg$^2$ and all 716 single-dishes sources  detected at $\geq$\,4.0-$\sigma$ significance were imaged at 870\,$\mu$m with ALMA as the ALMA SCUBA-2 UDS Survey (AS2UDS) \citep{simpson2015scuba,stach2018alma,stach2019alma}. This resulted in the largest, homogeneously-selected sample of SMGs to date across a contiguous field with excellent multi-wavelength coverage from which robust photometric redshifts could be derived both for the ALMA-detected SMGs and the $>$\,300,000  $K$-detected galaxies in this field \citep{dudzevivciute2020alma}. In this paper we present the results of the projected two-point cross-correlation analysis of the SMGs with the $K$-band detected field galaxy sample (Almaini et al.\ in prep.) utilising the full redshift PDFs from \cite{dudzevivciute2020alma} to constrain SMG clustering at redshifts $z$\,$=$\,1.5--3.0 free from the potential biases due to misidentifications, incompleteness and source blending that have undermined the conclusions from previous studies.

The structure of the paper is as follows. In \S\ref{sec:method} we describe the sample selection for the SMGs and the field galaxies used in our cross-correlation analysis and give a brief description of our method for measuring the clustering strengths of the AS2UDS SMGs. In \S\ref{sec:results} we present the results and discussion of our clustering analysis, including the dark matter halo masses as a function of redshift for our  sample and the comparisons with previous SMG clustering results. \S\ref{sec:conclusions} presents our main conclusions. Throughout this paper we assume a {\it Planck} XIII cosmology with $\Omega_{\rm m}$\,$=$\,0.307, H$_{0}$\,$=$\,69.3\,km\,s$^{-1}$\,Mpc$^{-1}$ (and using the standard definition for $h$ from H$_{0}$\,$=$\,100\,$h$\,km\,s$^{-1}$\,Mpc$^{-1}$), and for the amplitude of the matter power spectrum we use $\sigma_8$\,$=$\,0.816. All quoted magnitudes are on the AB system.

%
%
%
\section{Methodology} \label{sec:method}

\subsection{Sample Selection} \label{sec:sample}

Our clustering analysis employs a similar cross-correlation method as used by \cite{hickox2012laboca}. 
We focus on this methodology as it allows the inclusion of spectroscopic and photometric redshift information in the clustering analysis and hence it is likely to be increasingly adopted in future studies as the available redshift information expands on submillimetre galaxy samples.
We therefore  start by defining four catalogues: an SMG catalogue (`SMGs'), a comparison population within the same volume as the SMGs (`Galaxies'), and randomised distributions for both SMGs and the comparison sample (`Random'). For the SMG catalogue we use as a basis the 707 SMGs in the  catalogue from \cite{stach2019alma}'s
AS2UDS ALMA survey in the UKIRT Infrared Deep Sky Survey  \citep[UKIDSS,][]{lawrence2007ukirt} Ultra-Deep Survey field (UDS, Almaini et al.\ in prep.), which we briefly discuss here \citep[for a full description see:][]{stach2019alma,dudzevivciute2020alma}. The AS2UDS survey obtained ALMA Band 7 (870$\mu$m) continuum observations of all $>$\,4-$\sigma$ sources from the S2CLS SCUBA-2 850-$\mu$m map of the UDS region \citep{geach2017scuba}. This SCUBA-2 map also  formed the basis for the earlier \cite{wilkinson2016scuba} clustering analysis, that relied upon  probabilistically-identified radio, mid-infrared and colour-selected counterparts to the single-dish submillimetre sources (some of which were subsequently shown to be incorrect, see \citet{an2018machine}).  In contrast, we now  have robust ALMA interferometric identifications, at 870\,$\mu$m and $\sim$\,0.3$''$ resolution,  of the true counterparts to the single-dish sources. The  S2CLS UDS map reached a median depth of 0.9\,mJy\,beam$^{-1}$ across the 0.96\,deg$^{2}$ \citep{geach2017scuba}.
ALMA continuum mapping of  716 single-dish sources  located 708 submillimetre galaxies  at $>$\,4.3-$\sigma$ significance (spanning a flux range of $S_{870}$\,$=$\,0.6--13.6\,mJy), with this selection threshold corresponding to a 2 per cent false-positive rate. 
We  note that, as described in \cite{stach2019alma}, to remove any bias against detecting  extended sources, the sources were detected from ALMA continuum maps which were $uv$-tapered to 0.5$''$  resolution.
In addition,  previously discovered very bright, $z$\,$=$\,3.4 strongly lensed source \citep{ikarashi2011detection} was removed from our SMG catalogue resulting in 707 SMGs in this catalogue \citep{stach2019alma}. 

The properties of the ALMA source catalogue from the AS2UDS survey are described in \cite{stach2018alma,stach2019alma}.  These papers include analysis illustrating the recovered fraction of the
single-dish flux density arising from detected ALMA components, as well as 
variation of median redshift with submillimetre flux density for the ALMA-detected galaxies \citep[both included in][]{stach2019alma}, as well as the trends in multiplicity of the single-dish SCUBA-2 sources with flux density \citep{stach2018alma}.  We note in particular that \cite{stach2018alma} conclude that the majority of SCUBA-2 sources with multiple ALMA-detected components arise due to the projection of faint, unrelated submillimetre galaxies in the vicinity of brighter sources, rather than due to intrinsically clustered sources, a point we return to in our analysis.  Finally, to assess the influence of any inhomogeneity in the properties of the original SCUBA-2 catalogue which was the basis for our ALMA follow-up observations, we also investigate the effect of applying a $S_{870\mu m}$\,$\geq$\,4.0\,mJy flux density (corresponding to 4.4-$\sigma$ in the SCUBA-2 map) cut on our ALMA sample, as the parent single-dish catalogue is expected to
be close to $\sim$\,100 per cent complete for sources brighter than this limit.

%
%
\begin{figure}
 \includegraphics[width=\columnwidth]{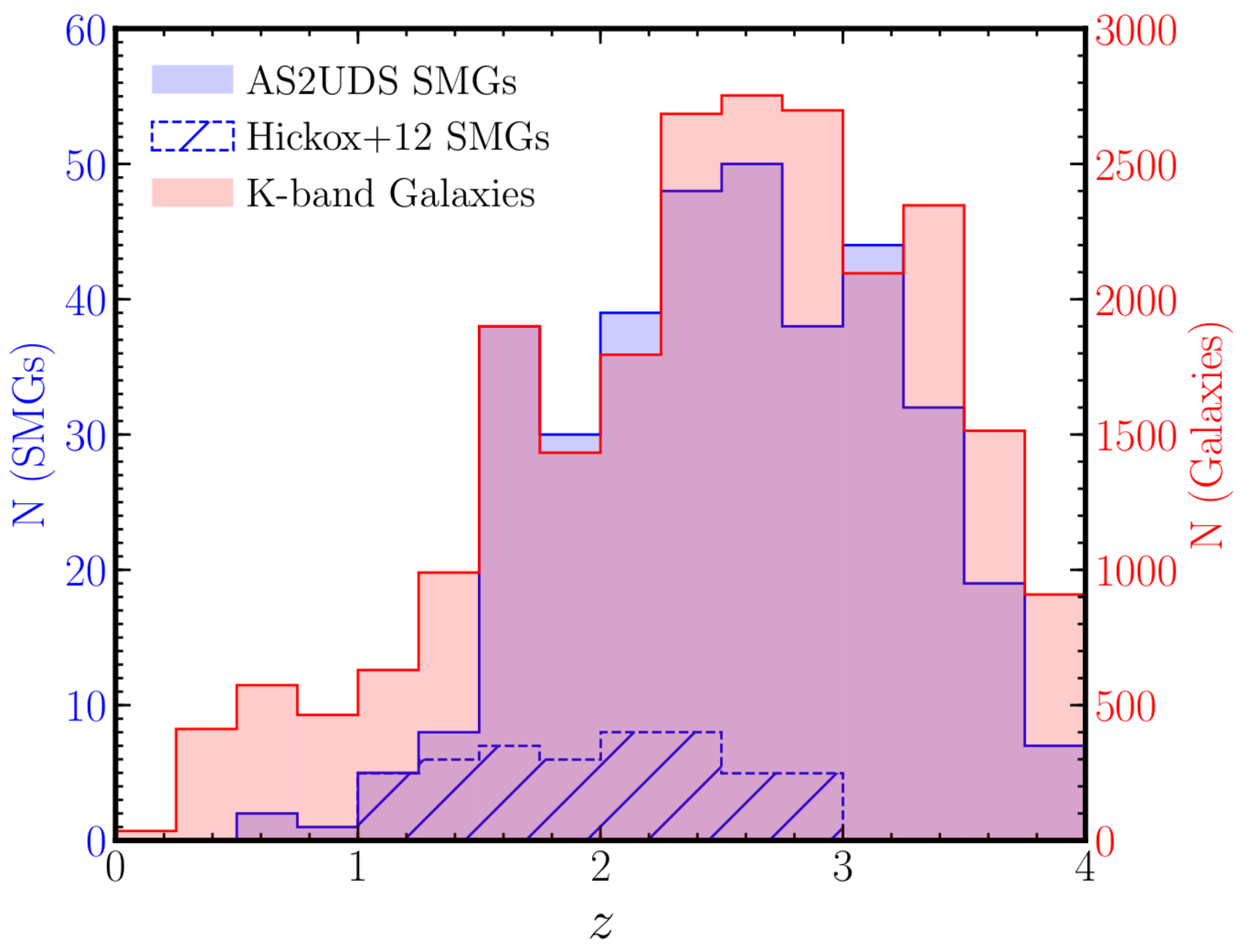}
 \caption{The redshift distributions for the SMG sample from the AS2UDS survey in comparison to the  probabilistically-identified SMGs in the same redshift range from the LESS survey analysed by \protect\cite{hickox2012laboca} (hatched histogram). This illustrates the $\sim$\,7$\times$ increase in sample size that comes from our
 analysis of the deep $\sim$\,1\,deg$^2$ UDS field compared to the somewhat shallower $\sim$\,0.25\,deg$^2$ LESS survey. In addition, our analysis benefits from robust and unambiguous ALMA-identified counterparts, compared to the
 earlier study. The redshifts for the AS2UDS sample are the photometric redshifts reported for these ALMA-identified SMGs from \protect\cite{dudzevivciute2020alma}, whilst the LESS sample are a combination of spectroscopic redshifts and photometric redshifts from \protect\cite{wardlow2011laboca} for the probabilistic radio/mid-infrared-identified counterparts \protect\citep[subsequent ALMA studies are reported by][]{hodge2013alma,danielson2017alma}. Also shown is the redshift distribution for the $K$-selected field galaxies that have been redshift-matched to the SMG sample as discussed in  \S\ref{sec:sample}, that are used for the projected cross-correlation analysis with the SMGs, the scale for this sample is shown on the right-hand ordinate.}
 \label{fig:Fig1}
\end{figure}

\subsection{Photometric Redshifts}

Photometric redshifts and other physical parameters (as well as their associated probability density functions) were estimated by \cite{dudzevivciute2020alma} through spectral energy distribution (SED) fitting of  the multi-wavelength ultra-violet--to--radio coverage in the UDS field using the \textsc{magphys} modelling code \citep{da2008simple,da2015alma,battisti2019magphys+}. For full details of the results from the \textsc{magphys} analysis and the extensive testing of these, see \cite{dudzevivciute2020alma}. Here we provide a brief description of the testing of the photometric redshifts that is relevant to this work. 

The uncertainties on the redshifts (and thus the broadness of their PDFs) is reliant in part on the number of constraints to the SED, i.e.\ the number of photometric bands with detections or limits for each galaxy. For the subset of optically-bright SMGs with detections in all 22 photometric bands available, the \textsc{magphys} PDFs are narrow with a 16--84\,th percentile range of $\Delta z$\,$\sim$\,0.2 but this broadens to $\Delta z$\,$\sim$\,0.5 for SMGs detected in only 12 photometric bands. 
The absolute accuracy of the \textsc{magphys} redshifts were tested in \cite{dudzevivciute2020alma} by comparing \textsc{magphys}-derived photometric redshifts with existing spectroscopic redshifts \citep[e.g.][Hartley et al.\ in prep.; Almaini et al.\ in prep.]{smail2008pilot} from both 6,719 field galaxies which gave a median photometric offset of $\Delta z$/(1+$z_{\rm spec})$\,$=$\,$-$0.005$\pm$0.003 and for a subset of 44 SMGs with  spectroscopic redshifts at a median offset of $\Delta z$/(1+$z_{\rm spec})$\,$=$\,$-$0.02$\pm$0.03 (we also employ these spectroscopic redshifts in our analysis for the small number of SMGs for which they are available). Similarly in \cite{birkin2020alma} the accuracy of the AS2UDS \textsc{magphys} PDFs were tested against newly acquired and unambiguous CO-derived millimetre spectroscopic redshifts of a sample of 16 SMGs with the PDFs correctly identifying the SMG redshift for 14 of the cases (88 per cent). 

\begin{figure}
 \includegraphics[width=\columnwidth]{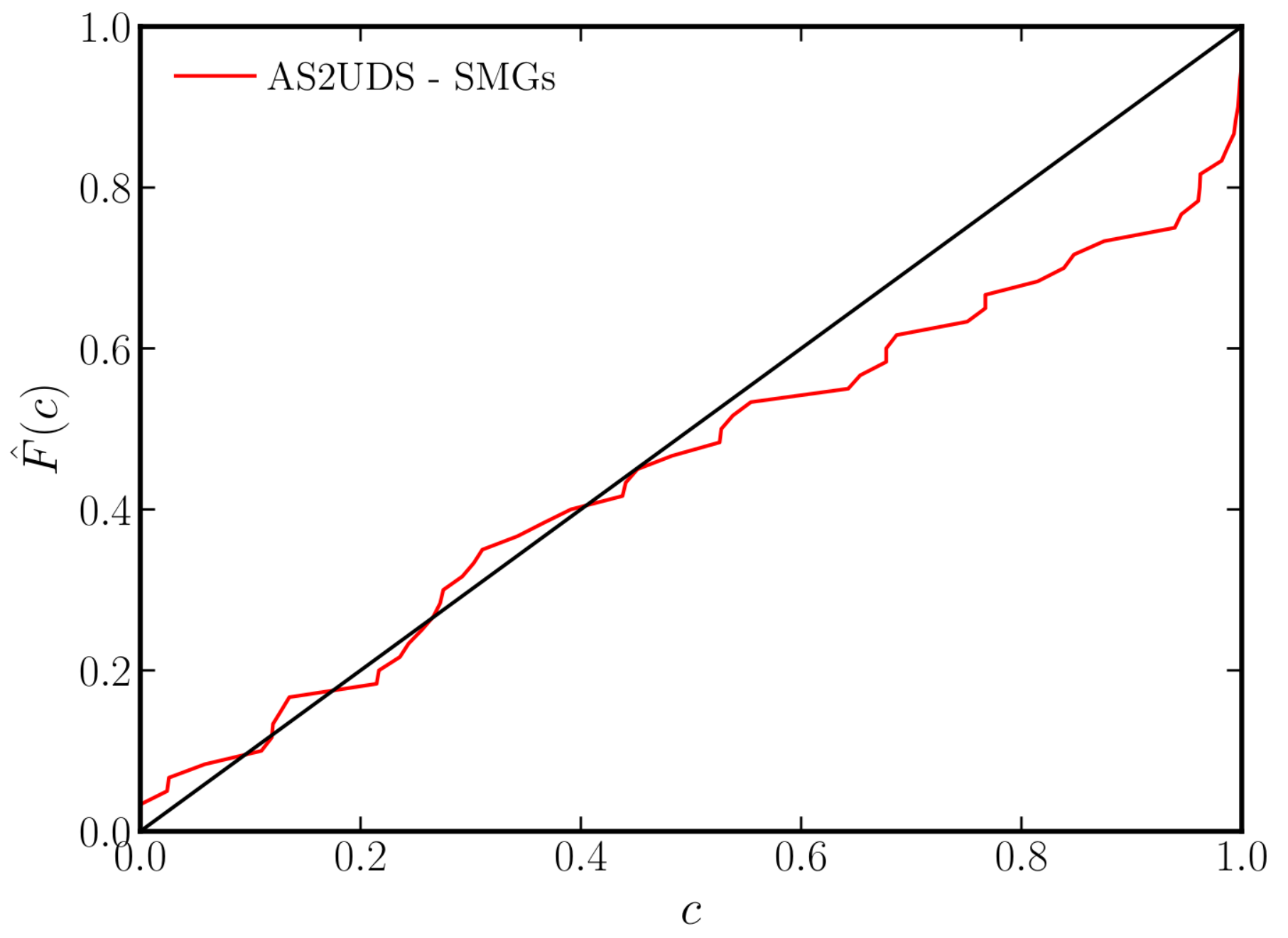}
 \caption{The empirical cumulative distribution function of the threshold credibilities for those AS2UDS SMGs with spectroscopic redshifts. The interpretation of the trend in this  plot is that a sample that falls below the one-to-one relation, shown in black, results  from photometric redshift PDFs that are on average narrower than expected given their spectroscopic to photometric redshifts offsets, whilst samples with lines above the one-to-one relation have overly broader PDFs and are thus on average overestimating the photometric redshift uncertainties. For our \textsc{magphys}  PDFs we find that the trend is consistent with the one-to-one relation and hence that the PDFs appear to accurately reflect the uncertainties expected from the observed offsets of the  photometric redshifts compared to the available spectroscopic redshifts.}
 \label{fig:ztest}
\end{figure}

Our Monte-Carlo method described below is sensitive to not just the accuracy of the median photometric redshift offset but also the accuracy of the redshift PDF. We therefore tested the accuracy of the \textsc{magphys} PDFs using the method outlined in \cite{wittman2016overconfidence}, where for the SMGs where we have spectroscopic redshifts ($z_s$) we measure how many lie within their expected confidence interval of their respective predicted photometric redshift PDFs, i.e. do 1 per cent of the $z_s$ lie within the 1 per cent confidence interval, 10 per cent within the 10 per cent confidence interval, etc. This is achieved by measuring the fraction of the PDF that lies within the redshift intervals where the distribution is greater than the value of the PDF at the spectroscopic redshift $p(z_s)$, i.e.:

\begin{equation}
c_{i}=\sum_{z \in p_{i}(z) \geq p_{i}\left(z_{s, i}\right)} p_{i}(z)
\end{equation}

where $p_{i}(z)$ is the PDF for the $i$th SMG with a spectroscopic redshift and $c_i$ is the resulting threshold credibility. The empirical cumulative distribution function of these threshold credibilities, $\hat{F}(c)$, should then follow a one-to-one relation with $c$ if the redshift PDFs were accurately measuring the uncertainties in the photometric redshifts. If the cumulative distribution falls below the unity relation then it suggests the PDFs are underestimating the uncertainties (the peaks are too narrow) and likewise if the distribution is above the line then this suggests the PDFs are over-estimating the uncertainties. In Figure \ref{fig:ztest} we show the cumulative distribution plot for our AS2UDS SMGs which shows the hint of an underestimation in the redshift uncertainties but a Kolmogorov-Smirnov test of the SMGs against the ideal one-to-one relation finds a probability of this occuring by chance of 12 per cent suggesting that this is not a statistically significant deviation. Therefore, from the  sample of SMGs for which we have spectroscopic redshifts, we conclude that \text{magphys} returns both accurate photometric redshifts and representative uncertainties. In addition, we tested the potential impact of an underestimation or overestimation of the broadness of the \textsc{magphys} derived redshift PDFs by replacing the PDFs of sources in our analysis with Gaussians of varying widths, greater and smaller than $\Delta z$\,$\sim$\,0.5, centred at the photometric redshifts. We find that varying the widths of the PDFs in this way has a minimal impact on the resulting cross-correlation functions but does impact the uncertainties, with broader PDFs resulting in larger uncertainties.

%
%

\subsection{\emph{K}-band Galaxies}

For the cross-correlation analysis we make use of the UKIDSS UDS DR11 catalogue (Almaini et al.\ in prep.), a $K$-band selected catalogue that covers the majority of the S2CLS map of the UDS field (634/707 SMGs covered) which has been matched to up-to 21 other photometric bands from the ultra-violet to radio \citep[these same photometric data are used to model the SMG sample, see][]{an2018machine,dudzevivciute2020alma}. The DR11 catalogue contains 296,007 sources extracted from the $K$-band image with a median 5-$\sigma$ depth of $K$\,$=$\,25.3\,mag, however close to a third of the UKIDSS coverage is flagged due to no optical coverage and thus insufficient photometry for `good' quality photometric redshifts for the comparison galaxy sample. We mask these regions from all clustering input catalogues, i.e.\ the SMGs, Galaxies and Randoms. The photometric redshifts for the remaining galaxies and their associated PDFs were estimated by \cite{dudzevivciute2020alma} in an analogous manner to the SMGs.

As mentioned above we created a `mask' that flagged those regions within the UDS field that were either not covered by the UDS-DR11 catalogue, not covered by the S2CLS SCUBA-2 map, or were flagged as potentially contaminated photometry in the UDS-DR11 coverage. This mask was applied to both the UDS-DR11 $K$-band `Galaxy' catalogue and the AS2UDS `SMG' catalogue and the  two associated `Random' catalogues. These `Random' catalogues were made by randomly assigning spatial positions for galaxies within these unmasked regions at approximately ten times the density of SMGs to create the Random$_{\rm SMG}$ catalogue and, again, approximately ten times the density of the $K$-band galaxies to form a Random$_{\rm Gal}$. Then each source in these catalogues was assigned a redshift by sampling the associated mean photometric redshift PDFs for the `real' SMGs and galaxies.

To better define our `Galaxy' sample we first note that the fraction of the $K$-band selected galaxy sample at redshifts $z$\,$>$\,3 decreases rapidly in comparison to our SMG sample, and similarly the SMG sample has comparatively few sources at $z$\,$<$\,1,  thus to ensure a statistically robust Galaxy sample to cross-correlate with our SMGs  we restrict our clustering measurements to $z$\,$=$\,1.5--3.0, thus maximising the overlap in the redshift distributions between the SMGs and $K$-selected galaxies (Figure~\ref{fig:Fig1}). Next, 
we follow previous clustering studies in the UDS field \citep{hartley2013studying,wilkinson2016scuba} and apply a 90 per cent mass completeness limit using our new \textsc{magphys} masses. These previous  studies have applied a redshift dependant mass limit, but from our experimentation the resulting clustering results are insensitive to the minor evolution in the mass completeness limits with redshift across the redshift
range of interest.  We therefore apply a uniform 90 per cent mass completeness limit for our $z$\,$=$\,1.5--3 comparison galaxies of $M_{\rm lim}$>$10^{9.1}$\,M$_{\odot}$ to all redshift bins, removing the lowest stellar mass galaxies that have no analogues in the SMG sample. 
Finally, because the redshift distribution for the mass-limited galaxy sample peaks at a significantly lower redshift ($z$\,$=$\,1.93) than the SMGs ($z$\,$=$\,2.61$\pm$0.08) our SMG--Galaxy relative bias measurement will be dominated by more luminous, higher-redshift galaxies. Therefore we maximise the cross-correlation signal by randomly selecting galaxies from the $K$-band galaxy catalogue such that their resulting  redshift distribution approximately matches the distribution of the SMGs (Figure~\ref{fig:Fig1}). 

\subsection{Projected Cross-Correlations}

To derive the halo masses of our sample of SMGs we have to estimate their two-point correlation function $\xi(r)$. This function is defined as the probability $P$ above Poisson of finding two galaxies physically separated a distance $r$ in a volume element $dV$, i.e.:
\begin{equation}
d P=n[1+\xi(r)] d V,
\end{equation}
where $n$ is the mean space density of the galaxies. In the projected two-point correlation function we project the $r$ separation into two components, perpendicular ($r_{p}$) and parallel ($\pi$) to the  line-of-sight. The projected correlation function $w_{p}(r_{p})$ is then defined as the integral of the correlation function, $\xi(r)$,  over the line-of-sight:
\begin{equation}
w_{p}(r_{p})=2 \int_{0}^{\pi_{\max }} \xi(r_{p}, \pi) d \pi.
\end{equation}
Following \cite{davis1983survey} we measure the line-of-sight separations from the co-moving radial distances, $D$, derived from their redshifts ($\pi_{21}$\,$=$\,$D_2-D_1$). The perpendicular components for each galaxy pair can then be calculated from the on-sky separations ($\theta$) and radial co-moving distances simply using the cosine rule:
\begin{equation}
    r_{p}=[2D_2D_1(1-\cos{\theta})]^{1/2}
\end{equation}

By integrating the correlation function by a suitable distance along the line-of-sight the issues of redshift space distortions \citep{kaiser1987clustering} owing to the peculiar velocities of the galaxies are removed. $\xi(r)$ can be approximated by a simple power-law in the form $\xi(r)=(r/r_0)^{-\gamma}$ and we choose to fix $\gamma$\,$=$\,1.8 which is a value consistent with both our measurements, where we allow the slope to vary in our fits, and with that found from many previous studies of galaxies and SMGs \citep[e.g.][]{zehavi2005luminosity,farrah2006spatial,coil2007deep2,hickox2012laboca}, and also has been chosen by the majority of the previous literature results that we compare to. 

If we adopt a power-law parameterisation of the real space correlation function $\xi(r)$ then, from \cite{peebles1980large}, this can be related to the projected cross correlation function $w_{p}(r_{p}$) using:
\begin{equation} \label{eq:powerlaw}
w_{p}(r_p)=r_p\left(\frac{r_{0}}{r_p}\right)^{\gamma} \frac{\Gamma(1 / 2) \Gamma[(\gamma-1) / 2]}{\Gamma(\gamma / 2)},
\end{equation}
\smallskip
where $\Gamma$ is the gamma function. Therefore from a simple power-law fit to the projected correlation function we can directly estimate the correlation length $r_0$ for the corresponding galaxies. The power-law parameterisation does assume we integrate to $\pi_{max}$\,$=$\,$\infty$, however the integral in Equation~2 is in practice limited to a set co-moving distance which has to be large enough to recover all the clustering signal, but small enough to reduce the noise from including uncorrelated pairs at larger separations. For this work we chose a value of $\pi_{max}$\,$=$\,100\,$h^{-1}$\,Mpc that is consistent with \cite{hickox2011clustering,hickox2012laboca} who also fitted projected cross-correlation functions using photometric redshifts and their PDFs in a similar redshift range to this work. There are however other studies with projected correlation functions using photometric redshifts which apply larger $\pi_{max}$ values of $\sim$\,400 Mpc\,h$^{-1}$ \citep{georgakakis2014large}, to try and encapsulate the larger uncertainties inherent with photometric redshifts. To test the impact of increasing $\pi_{max}$ we estimated $r_0$ values for our full redshift sample described below with the increased $\pi_{max}$\,$=$\,400\,Mpc\,h$^{-1}$. We found that the $r_0$ value was largely insensitive to this increase, showing only an $\sim$\,10 per cent increase in $r_0$ from $\pi_{max}$\,$=$\,100\,Mpc\,h$^{-1}$, which is well encapsulated by the substantial uncertainties in our derived $r_0$ values, thus we retain  $\pi_{max}$\,$=$\,100\,Mpc\,h$^{-1}$ for consistency with \cite{hickox2012laboca}.

To estimate the correlation function we use the \cite{landy1993bias} estimator for cross-correlation:
\begin{equation}
\xi\left(r_{p}, \pi\right)=\frac{D_{\rm SMG} D_{\rm Gal}-D_{\rm SMG} R_{\rm SMG}-R_{\rm Gal} D_{\rm Gal}+R_{\rm SMG} R_{\rm Gal}}{R_{\rm SMG} R_{\rm Gal}}
\end{equation}
where $D_{\rm SMG}D_{\rm Gal}$ is the normalised number of SMG--Galaxy pairs, $D_{\rm SMG}R_{\rm SMG}$ SMG--Random$_{\rm SMG}$, $R_{\rm Gal}D_{\rm Gal}$ Galaxy--Random$_{\rm Gal}$ and $R_{\rm SMG}R_{\rm Gal}$ the Random$_{\rm SMG}$--Random$_{\rm Gal}$ pairs at separations $r_{p}\pm\Delta r_{p}$ and $\pi\pm\Delta\pi$. For our cross-correlation analysis we calculated pair counts in logarithmic $r_p$ bins in the range 0.05--14\,$h^{-1}$\,Mpc, which at the median redshift of the SMG sample ($z$\,$\sim$\,2.5) corresponds to angular scales in the range $\sim$\,2.5--700$''$.

To incorporate the photometric redshift PDFs we measure the projected correlation function with a Monte Carlo method by repeating the projected correlation as a function of $r_p$ bins whilst sampling the redshifts of every SMG and galaxy by randomly selecting from their respective PDFs. We set the contribution to the uncertainties for the final estimation of the projected correlation function from the sampling as the 16th and 84th percentile of the $w_p(r_p)$ distribution in each $r_p$ bin from the resulting 3,000 redshift-sampling iterations which are combined with the poisson uncertainties estimated from the median pair counts for each bin. For the small subset of SMGs with archival spectroscopic redshifts (37 SMGs with $z$\,$=$\,1.5--3) the PDFs were set to delta functions at the spectroscopic redshifts.

\subsection{Galaxy--Galaxy Auto-correlation}

The projected cross-correlation of the SMGs with the $K$-band field galaxies provides us with the \textit{relative} bias between SMGs and these galaxies as described below. However, to estimate the characteristic halo mass of the SMGs we need to determine the absolute bias of the SMGs relative to the dark matter and thus we need to estimate the absolute bias of the galaxies with respect to the dark matter. To determine this we measure the auto-correlation function for the comparison $K$-band galaxies that were redshift-matched to the SMGs. This redshift-matched galaxy sample  is  large enough ($N$\,$\sim$\,50,000) that it is possible to measure a signal from its auto-correlation function. In addition, to reduce any uncertainty in the auto-correlation function due to the individual photometric redshifts of these galaxies, we measure an \textit{angular} auto-correlation function for this sample. We  measure the angular auto-correlation using the same \cite{landy1993bias} estimator, but now modified for auto-correlation:
\begin{equation}
\omega(\theta)=\frac{1}{R R}(D D-2 D R+R R)
\end{equation}
where $DD$, $DR$, and $RR$ are the normalised number of Galaxy--Galaxy, Galaxy--Random, and Random--Random galaxy pairs, respectively, at an angular separation $\theta$. The errors for the auto-correlation function are calculated by dividing the field into nine roughly equal sized sub-fields and using the `delete one jackknife' method \citep{shao1986discussion,norberg2009statistical}.

As with the projected correlation function, we fit a power-law to the galaxy auto-correlation function of the form $w(\theta)$\,$=$\,$A\theta^{-\delta}$. To measure the absolute correlation length for the SMGs we convert $A$ and $\delta$ to the projected correlation function equivalent $r_0$ and $\gamma$ following \cite{peebles1980large} by deprojecting the auto-correlation function through the Limber equation \citep{limber1954analysis}:
\begin{equation}
    \delta=\gamma-1,
\end{equation}
and 
\begin{equation}
A=H_{\gamma} \frac{\int_{0}^{\infty}\left(d N_{1} / d z\right)\left(d N_{2} / d z\right) E_{z} \chi^{1-\gamma} d z}{\left[\int_{0}^{\infty}\left(d N_{1} / d z\right) d z\right]\left[\int_{0}^{\infty}\left(d N_{2} / d z\right) d z\right]} r_{0}^{\gamma}
\end{equation}
where 
\begin{equation}
H_{\gamma}=\frac{\Gamma(0.5) \Gamma(0.5[\gamma-1])}{\Gamma(0.5 \gamma)},
\end{equation}
and $dN_1/dz$ and $dN_2/dz$ are the redshift distribution for the samples, where for our auto-correlation $dN_1/dz=dN_2/dz$ and $E_z$ is $H_z/c$ where $H_z$ is the Hubble parameter. With an $r_0$ for the SMG--Galaxy cross-correlation and the Galaxy--Galaxy auto-correlation we can estimate the correlation length for the auto-correlation function of SMGs, using a fixed $\gamma$\,$=$\,1.8 for the three correlation functions, from $\xi_{\rm SMG}$\,$=$\,$\xi_{\rm SMG-Gal}^2/\xi_{\rm Gal}$ \citep{coil2009aegis}.

\subsection{Deriving dark matter halo masses} \label{sec:dmhm}

As described above, to estimate the dark matter halo masses for the SMGs we first must measure the absolute bias of the $K$-band Galaxy sample from their auto-correlation. A measurement of bias relies on an estimation of the dark matter angular correlation function $w_{dm}$. To estimate $w_{dm}$ we  use the {\sc HaloFit} code \citep{smith2003stable}, with the updated {\sc Halofit} fitting parameters from \cite{takahashi2012revising}, to calculate the non-linear dark matter power spectrum $P(k,z)$ assuming the slope of the initial fluctuation power spectrum $\Gamma$\,$=$\,0.21. We use the \textsc{halomod} code (Murray et al.\ in prep.) to then project the power spectrum into the angular auto-correlation of the dark matter through Limber's equation \citep{limber1954analysis}:
\begin{equation}
w(\theta)=\frac{1}{c} \int\left(\frac{d N}{d z}\right)^{2} H(z) \int \frac{k}{2 \pi} P(k, z) J_{0}[k \theta\chi(z)] d k d z
\end{equation}
where $J_0$ is the zeroth order Bessel function, $\chi$ is the radial co-moving distance, and $dN/dz$ is the stacked redshift PDF normalised such that $\int_0^{\infty}[dN/dz]dz$\,$=$\,1.

Knowing the dark matter angular correlation function we can then measure the absolute bias ($b_g$) of the auto-correlated galaxies by scaling the dark matter angular correlation function to the galaxy correlation function:
\begin{equation}
w(\theta)=b_{gal}^{2} w_{d m}(\theta).
\end{equation}
For the relative SMG--Galaxy bias we calculate the dark matter projected correlation function from the same power spectrum, but now Fourier transformed to give $\xi(r)$ which is then integrated using Equation~3. 

The projected dark matter correlation function is scaled to the SMG--Galaxy projected cross-correlation function in the same manner as for the galaxy auto-correlation, but with the linear scaling now equal to $b_{\rm smg}b_{\rm gal}$ and thus the absolute SMG bias can be calculated. We convert the absolute bias into a dark matter halo mass by assuming a \cite{tinker2010large} bias function.

 From recent studies using Halo Occupation Distribution (HOD) models \citep{peacock2000halo,benson2000nature}, the galaxy correlation function can be broken down into the sum of two components, a `one-halo' term that dominates at smaller spatial scales and measures the contributions from pairs of galaxies within a single dark matter halo, and the `two-halo' term dominating at larger scales involving clustering of galaxies in separate haloes. Whilst we lack the signal-to-noise in our SMG correlation functions to constrain the increased number of parameters involved in fitting HOD models, we still test the potential influence of a contribution from a one-halo component on our results by 
 restricting the minimum spatial scale where we apply both the power-law and dark matter correlation function fits to minimise contributions from the intra-halo pairs. From the HOD fitting of 250\,$\mu$m selected SMGs in \cite{amvrosiadis2019herschel} we set the minimum spatial scale to $r_p>$\,0.5\,$h^{-1}$\,Mpc, this value chosen to be roughly consistent with the region in which the `one-halo' term dominates their galaxy clustering function.  We indicate this in Fig.~\ref{fig:Fig2} by showing the region used in the fitting as a solid line.

\subsection{Integral Constraint}

As our clustering measurements are for a sample in a finite-area field we check the magnitude of the correction to the measured correlation function from the absence of information on density fluctuations on the scale of the field  from the `integral constraint' (IC):
\begin{equation}
w_{p}=w_{p}^{\rm obs}+\text{IC}
\end{equation}
As we are constraining our fits to relatively small scales ($<$\,14\,$h^{-1}$\,Mpc)
in comparison to the degree-scale UDS field we expect this offset to be negligible in comparison to the measured clustering \citep[e.g.][]{kashino2017fmos}. We estimate the integral constraint for the projected correlation function using our `Random' catalogues by following the iterative method of \cite{roche1999angular} and \cite{kashino2017fmos}:
\begin{equation}
\text{IC}=2 \int_{0}^{\pi_{\max }} \frac{\sum_{i} R R\left(r_{i}\right) \xi^{\bmod }\left(r_{i}\right)}{\sum_{i} R R\left(r_{i}\right)} d \pi
\end{equation}
where $\xi^{\bmod }$ is the bias-scaled correlation function of the dark matter: $\xi^{\bmod }$\,$=$\,$b^2\xi_{\text dm}$ where $b$ is initially set to the SMG--Galaxy relative bias as measured in \S\ref{sec:dmhm}. The integral constraint is then calculated and applied and the process is repeated with the updated relative bias values incorporating the estimated integral constraint offsets until convergence. We find a final integral constraint correction for our field of $\rm{IC}$\,$=$\,2.0\,$h^{-1}$\,Mpc, which is an insignificant correction to the observed correlation function at the separation scales we consider. Nevertheless, we still correct our observed correlation function amplitudes for this integral constraint. 
The galaxy auto-correlation amplitudes were similarly corrected for the integral constraint, estimated with: 
\begin{equation}
\text{IC}=\frac{\sum_{i} R R\left(\theta_{i}\right) w\left(\theta_{i}\right)}{\sum_{i} R R\left(\theta_{i}\right)}
\end{equation}
where, following \cite{hartley2013studying}, for $w(\theta)$ we use the angular correlation function of the dark matter correlation function traced by our galaxy sample, scaled by the absolute galaxy bias, as measured in \S\ref{sec:dmhm}.

%
%
%
\section{Results and Discussion} \label{sec:results}

There is a complexity in measuring the clustering of SMGs that are detected from follow-up surveys of single-dish observations, arising from a potential bias due to the low resolution of the parent survey, e.g.\ `blending bias' \cite{cowley2016clustering}, that could increase the measured clustering. These issues arise because the low-resolution single-dish observations not only detect individual galaxies, but also can uncover groups of faint SMGs (either physically associated or simply seen in projection) that are separated on the sky by less than the single-dish beam \citep[$\sim$30\,\% for S2CLS sources:][]{stach2018alma}. As expected, below the flux limit of the single-dish parent survey the ALMA follow-up survey is incomplete to fainter SMGs, but some faint galaxies at these flux limits are included due to such blending (the bulk arise due to noise-boosting), including those in true SMG groups, whose summed flux density raise them above the single-dish flux density detection threshold. Missing isolated examples of such  faint galaxies, but detecting those preferentially lying in small separation groups could result in an overestimation of the true SMG clustering. In addition, if angular correlation functions are used to derive the clustering measurements, then these groups of SMGs, even if just projected systems with a wide spread in redshift between the components, rather than physically associated, can become correlated if the redshift bins are coarse enough \citep{cowley2016clustering}. 

A recent test of the potential significance of blending bias was shown by \cite{garcia2020clustering} who assessed the strength of the effect for the ALMA follow-up of the single-dish LABOCA sources used in \cite{hickox2012laboca},  by applying a complex forward modelling technique to attempt to both account for incompleteness and assess the clustering of the SMGs.  They suggested that such an approach 
is necessary to correctly return the true characteristic halo masses and that by just measuring clustering from the single-dish sources alone can result in a halo mass 3.8$^{+3.8}_{-2.6}$ times higher than the true mass. We expect that our sample is less sensitive to this bias than that used in the \cite{garcia2020clustering} analysis, as the AS2UDS SMGs are follow-up of SCUBA-2 single dish observations with SCUBA-2 having a smaller beam size than the beam-convolved LABOCA map on which their analysis is based (14.6$''$, c.f $\sim$\,25$''$), as well as the higher significance cut used as the basis for the subsequent ALMA follow-up observations (4.0\,$\sigma$, c.f 3.75\,$\sigma$). We also note that for our AS2UDS survey, of the 440 SCUBA-2 sources lying within the unmasked regions described in \S\ref{sec:sample}, only 57 of these SCUBA-2 sources contain more than a single SMG and thus are potentially introducing some `blending bias'. As only $\sim$\,13 per cent of our SCUBA-2 sources being considered here contain multiple SMGs (whether physically associated or not), and of those we only expect $\sim$\,30 per cent to be physically associated \citep{stach2018alma,simpson2020alma}, we therefore do not expect a significant overestimation of the clustering from these biases. Nevertheless, with our increased sample size we can test for the magnitude of the blending bias (and other inhomogeneities in the parent SCUBA-2 catalogue) by measuring the clustering for both the `full' SMG sample and for a subset of SMGs by applying a  cut at the flux density where we are close to 100 per cent complete in the single-dish survey (ALMA $S_{870}$\,$\geq$\,4.0\,mJy flux cut) thus removing both the lower significance sources and any groups of faint SMGs within a single-dish beam.

%
%
\begin{figure}
 \includegraphics[width=\columnwidth]{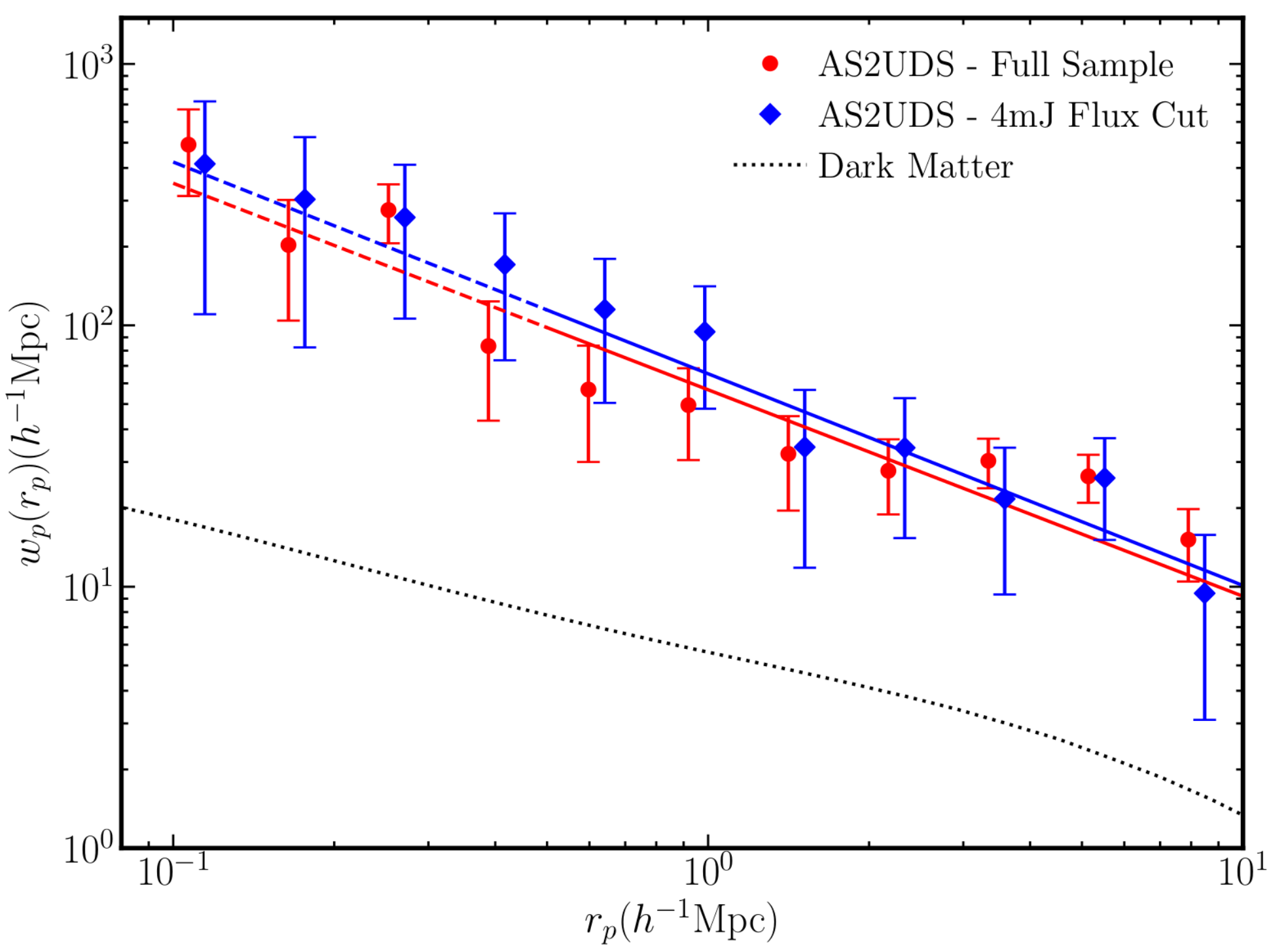}
 \caption{The projected cross-correlation function for the AS2UDS SMGs across the redshift range $z$\,$=$\,1.5--3.0, for the `full' sample and the subset with $S_{870}$\,$\geq$\,4.0\,mJy, which show no significant difference in the clustering amplitudes derived from the latter selection.  To quantify this we show single parameter power-law fits to the two samples by the dashed lines (plotted as solid in the region
 at $>$\,0.5\,$h^{-1}$\,Mpc used in the fitting), for which we derive correlation lengths of $r_0$\,$=$\,6.6$^{+1.9}_{-1.9}$\,$h^{-1}$\,Mpc for the `full' and $r_0$\,$=$\,7.7$^{+2.8}_{-2.6}$\,$h^{-1}$\,Mpc for the $S_{870}$\,$\geq$\,4.0-mJy  sub-sample (data points offset in $r_0$ for clarity). This confirms that the clustering measured from the two samples are statistically indistinguishable.  The dotted line shows the projected correlation of the underlying dark matter which is then linearly scaled to the samples to derive their relative bias measurements.}
 \label{fig:Fig2}
\end{figure}

In Figure~\ref{fig:Fig2} we show the projected cross-correlation function for the AS2UDS SMGs with the  $K$-band galaxies, both with and without the ALMA 870-$\mu$m flux cut mentioned above, across the redshift range of $z$\,$=$\,1.5--3, that is similar to the redshift ranges used in many of the earlier SMG clustering studies \citep[e.g.][]{blain2004clustering,hickox2012laboca,wilkinson2016scuba}. 
We  fit both cross-correlation functions on scales larger than 0.5\,$h^{-1}$\,Mpc (to reduce the influence of the one-halo term), using a maximum likelihood estimator with a single parameter, power-law model given in Equation~\ref{eq:powerlaw}, where we fix $\gamma$\,$=$\,1.8. This returns the cross-correlation  lengths that are then corrected for the auto-correlation length of the respective $K$-band galaxy samples to return the estimated SMG auto-correlation lengths $r_0$\,$=$\,6.6$^{+1.9}_{-1.9}$\,$h^{-1}$\,Mpc for the `full' sample and $r_0$\,$=$\,7.7$^{+2.8}_{-2.6}$\,$h^{-1}$\,Mpc for the $S_{870}$\,$\geq$\,4.0-mJy sub-sample. As can be seen there is  no significant difference in the average amplitude for the  correlation functions with and without the ALMA flux cut.
The small difference in correlation length, that is statistically insignificant, between the two samples is unsurprising as one might naively expect the lower flux SMGs would inhabit lower mass haloes and thus push the correlation length down. Although recent work at estimating the halo masses for the faintest SMGs ($S_{850}$\,$<$\,2\,mJy) suggests no variation in halo masses with submillimetre flux \citep{chen2016faint}. If there is no rapid variation in the halo mass for fainter SMGs, then these results suggest there is no significant influence for blending bias (or other inhomogeneities in the parent SCUBA-2 catalogue), that would arise from finding physically associated groups of faint SMGs as single-dish sources, on our clustering measurements.  Moreover, given we only expect to add $\sim$\,19 such SMGs to our  analysis if we do not apply the flux cut, this modest change is unsurprising and therefore for the remainder of the analysis we have therefore chosen to use the `full' sample.

%
%
\begin{table*}
 \caption{Clustering results for each redshift bin considered. The $z$\,$=$\,1.5--3.0  sample is for the `full' sample. Values in [] are the 2-$\sigma$ errors. }
 \label{tab:results}
 \begin{tabular}{ccccccc}
  \hline
  Redshift &  <$N_{\rm smg}$> & $r_0$ & $b_{\rm smg}$ & $\log_{10}(M_{\rm halo})$\\
   & & ($h^{-1}$\,Mpc) & & ($\log_{10}({h^{-1}\,\rm M_{\odot}})$) &  \\
  \hline
  1.5--3.0 & 329 & $6.6^{+1,9}_{-1.9}$& $4.1\pm0.7$ & $12.6^{+0.3[0.6]}_{-0.4[0.9]}$ \\
1.5--2.0 & 82 & $6.3^{+2.9}_{-2.7}$ & $3.0\pm0.7$ & $12.8^{+0.3[0.6]}_{-0.4[0.9]}$\\
2.0--2.5 & 108 & $6.1^{+2.7}_{-2.5}$ & $3.6\pm0.8$ & $12.6^{+0.3[0.6]}_{-0.4[1.0]}$\\
2.5--3.0 & 139 & $6.9^{+3.0}_{-2.7}$ & $4.8\pm0.9$ & $12.6^{+0.3[0.6]}_{-0.3[0.7]}$ \\
  \hline
 \end{tabular}
\end{table*}

%
%
\begin{figure*}
 \includegraphics[width=2\columnwidth]{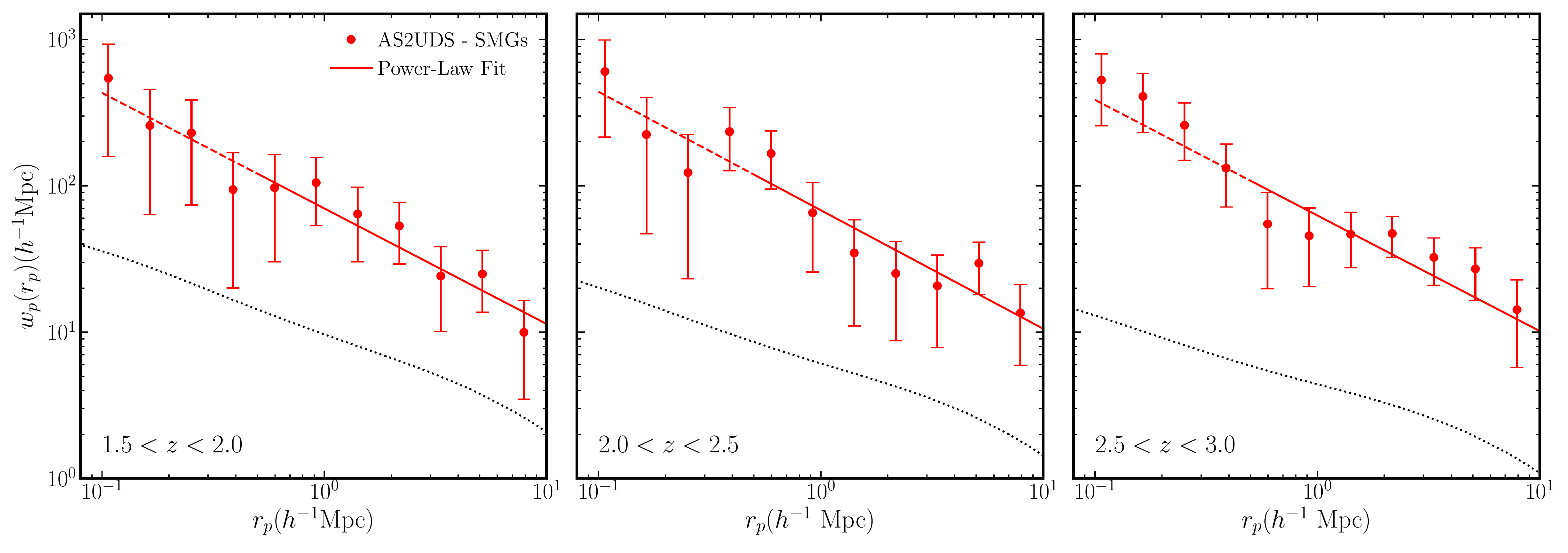}
 \caption{The two-point cross-correlation functions of submillimetre galaxies identified in the AS2UDS survey with redshift-matched $K$-band selected field galaxies from the UKIDSS UDS catalogue in three redshift bins ($z$\,$=$\,1.5--2.0, 2.0--2.5 and 2.5--3.0) giving  roughly equal number of SMGs per bin (Table~\ref{tab:results}). The solid lines are power-law fits to the cross-correlation with a fixed $\gamma$\,$=$\,1.8. The dotted lines show the projected auto-correlation of the dark matter. Combining the results of the power-law fits with the $K$-band galaxy auto-correlations, the dark matter halo masses for the submillimetre galaxies are derived and reported in Table~\ref{tab:results}).
 }
 \label{fig:Fig3}
\end{figure*}

\subsection{Correlation Length and Absolute Bias}

The correlation lengths are a useful measure for comparisons between different clustering studies as they are not dependent on different bias models, that can alter halo masses derived from the linear bias fitting. The weakest clustering (corresponding to the shortest correlation length) found for SMGs in the $z$\,$=$\,1.5--3 range was in \cite{wilkinson2016scuba} for probabilistically-identified counterparts to submillimetre sources in the same S2CLS map of the UDS region that is used as the basis of this study. They estimated a correlation length of $r_0$\,$=$\,4.1$^{+2.1}_{-2.0}$\,$h^{-1}$\,Mpc from their angular correlation functions which is $\sim$\,1\,$\sigma$ below our value for the same redshift range. As noted earlier,  \cite{wilkinson2016scuba} had to  rely on radio, mid-infrared and colour selection to identify likely counterparts
to the single-dish submillimetre sources, as opposed to high-resolution ALMA imaging.  Therefore contamination from mis-identifications \citep{hodge2013alma,an2018machine} is a likely cause for their lower correlation lengths; as they note, if they limit their analysis to the  more robust (but less complete) radio-identified counterparts they would estimate a longer correlation length: $r_0$\,$=$\,6.8$^{+2.7}_{-2.6}$\,$h^{-1}$\,Mpc, in better agreement with our measurements.

Comparing to clustering estimates in other fields from the literature, we find reasonable agreement with our measurements, e.g.\ in the Extended \textit{Chandra} Deep Field South, \cite{hickox2012laboca} also measured the projected correlation functions of probabalistically-identified counterparts to  single-dish detected sources and found a correlation length of $r_0$\,$=$\,7.7$^{+1.8}_{-2.3}$\,$h^{-1}$\,Mpc for sources at $z$\,$=$\,1--3, $r_0$\,$<$\,6--8\,$h^{-1}$\,Mpc in \cite{williams2011clustering} for SMGs selected at 1.1\,mm, and $r_0$\,$=$\,6.9$\pm$2.1\,$h^{-1}$\,Mpc in \cite{blain2004clustering}. All of these previous studies have suffered from modest sample sizes ($N$\,$<$\,100) and moreover their analysis were undertaken before large-scale, interferometric submillimetre  surveys were possible and so they  were often reliant on probabilistic multi-wavelength identifications, making them both incomplete for the higher redshift SMGs and also potentially contaminated from incorrect identifications.

%
%
\begin{figure*}
 \includegraphics[width=2\columnwidth]{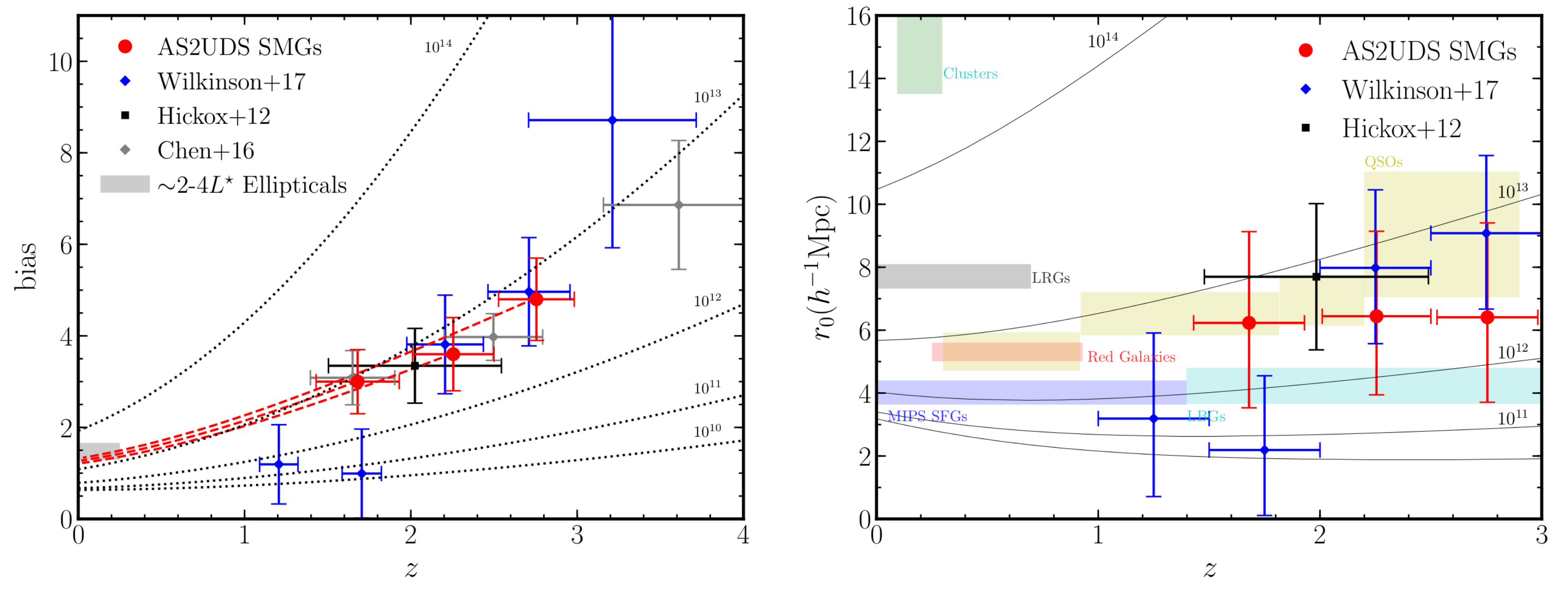}
 \caption{\textit{Left:} The predicted redshift evolution of the galaxy bias for the AS2UDS submillimetre galaxies.  The dotted lines show the expected bias evolution for dark matter haloes at their labelled masses. For comparison we show similar measurements from the \protect\cite{hickox2012laboca}, \protect\cite{chen2016scuba}  and \protect\cite{wilkinson2016scuba} SMG samples. Contrary to \protect\cite{wilkinson2016scuba}, but in agreement with \protect\cite{hickox2012laboca} and \protect\cite{chen2016scuba}, we do not see statistically significant evolution in the bias with redshift compared to that expected for a constant dark matter halo mass
 with a mass of $\sim$\,10$^{13}$\,M$_\odot$. The red dashed lines show the predicted halo mass growth rates from \protect\cite{fakhouri2010merger} for our three redshift bins.  These converge towards bias values for the descendant galaxy population at $z$\,$\sim$\,0 that are consistent from the bias' derived from the \protect\cite{zehavi2011galaxy} luminosity--bias relation for $\sim$\,2--4\,$L^{\star}$  galaxies, a population dominated by massive, passive spheroidal galaxies. \textit{Right:}  The auto-correlation lengths for our AS2UDS SMGs in three redshift bins compared to  observational estimates for a range of galaxy populations. The solid black curves show the expected correlation lengths for dark matter haloes of various masses. The AS2UDS SMGs at $z$\,$=$\,1.5--3 have longer correlation lengths than typically reported for
 UV-selected Lyman-break galaxies (LBGs), but similar to those
 measured for QSOs at comparable redshifts, suggesting that SMGs reside in haloes of similar mass to luminous QSOs.}
 \label{fig:Fig4}
\end{figure*}

Following \S\ref{sec:dmhm} we estimate the absolute bias for our `full'  sample finding $b_s$\,$=$\,4.1$\pm$0.7 (the $S_{870}$\,$\geq$\,4.0-mJy sample yields $b_s$\,$=$\,4.3$\pm$1.0), assuming a \cite{tinker2010large} bias model results in median halo masses of $\log_{10}(M_{\rm halo}[{h^{-1}\,\rm M_{\odot}}])$\,$=$\,12.6$^{+0.3}_{-0.4}$ for both samples. As with the correlation lengths, most previous studies are consistent with our estimates, e.g.\ $\log_{10}(M_{\rm halo}[{h^{-1}\,\rm M_{\odot}}])$\,$=$\,12.8$^{+0.3}_{-0.5}$ by \cite{hickox2012laboca}; $\log_{10}(M_{\rm halo}[{h^{-1}\,\rm M_{\odot}}])$\,$=$\,12.9$^{+0.2}_{-0.3}$ by \cite{chen2016faint}; and $\log_{10}(M_{\rm halo}[{h^{-1}\,\rm M_{\odot}}])$\,$\sim$\,12 from \cite{wilkinson2016scuba}.   Our direct estimate of the halo mass for SMGs in the AS2UDS survey also agrees well with that inferred from the redshift distribution of the AS2UDS SMG population by \cite{dudzevivciute2020alma}:  $\log_{10}(M_{\rm halo}[{h^{-1}\,\rm M_{\odot}}])$\,$\sim$\,12.8.  Their estimate is based on fitting the observed redshift distribution of SMGs with a model that combines an evolving gas fraction in haloes with a characteristic halo mass, which as it exceeded by a collapsing halo, leads to the formation of an SMG (this model was suggested by \citealt[][]{hickox2012laboca}, building on a similar model linking the clustering and redshift distribution of bright quasars in \citealt[][]{hickox2011clustering}). Our resulting halo mass however is above the proposed \textit{upper} limit for a $>$\,4.0\,mJy flux limited sample (the closest match from their results for our sample) of SMGs from \cite{garcia2020clustering} that suggests $\log_{10}(M_{\rm halo}[{h^{-1}\,\rm M_{\odot}}])$\,$<$\,12.22 but we can't rule out this limit as it lies within the 1-$\sigma$ error range of our characteristic mass. We return to discuss the connection between halo mass and redshift for the SMG population in \S\ref{sec:evolution} and \ref{sec:shmr}.

In comparison to theoretical simulations of SMGs, our median halo mass lies between the results from the semi-analytic model \textsc{galform} finding SMGs inhabiting haloes with masses $\log_{10}(M_{\rm halo}[{h^{-1}\,\rm M_{\odot}}])$\,$=$\,11.5--12 \citep{cowley2016clustering} and those from the N-body hydrodynamic simulation \textsc{eagle} that found that simulated SMGs with $S_{870}$\,$>$\,1\,mJy  reside in haloes with masses $\log_{10}(M_{\rm halo}[{h^{-1}\,\rm M_{\odot}}])$\,$=$\,12.96$^{+0.19}_{-0.01}$ \citep{mcalpine2019nature}.

\subsection{Clustering evolution with redshift} \label{sec:evolution}
\smallskip

We can exploit our relatively large sample size of $\sim$\,400 galaxies ($\sim$\,5--10$\times$ larger than similar previous studies) to split the sample into independent redshift bins to test any potential evolution in the halo masses. We split our sample into three redshift bins with equal $\Delta z$\,$=$\,0.5 yielding comparable numbers of SMGs in each (Table~\ref{tab:results}): $z$\,$=$\,1.5--2.0, $z$\,$=$\,2.0--2.5 and $z$\,$=$\,2.5--3.0.  We then repeat the clustering analysis as described above and we show the SMG--Galaxy cross-correlation functions for the three redshift bins in Figure~\ref{fig:Fig3}.

As with the single redshift bin, we derive the absolute bias for the SMGs residing in each redshift bin and the inferred halo masses. We show these in Figure~\ref{fig:Fig4} and list the associated bias values and halo masses in Table~\ref{tab:results}. In addition, we estimate the correlation lengths for our SMGs from the single-parameter power-law fits for each redshift bin; these are also reported in Table \ref{tab:results} and are shown in Figure \ref{fig:Fig4}. As with the single redshift bin, our correlation lengths lie slightly below the single measurement from \cite{hickox2012laboca} and are straddled by the results reported by \cite{wilkinson2016scuba}, although they are consistent within the relatively large uncertainties.  Overall, we see that all three redshift ranges show very similar correlation lengths, $r_0$\,$\sim$\,6--7\,$h^{-1}$\,Mpc, and inferred halo masses,  $\log_{10}(M_{\rm cen}[{\rm M_{\odot}}])$\,$\sim$\,12.6--12.8. The error values in parentheses for the halo masses represent the 2-$\sigma$ uncertainties.

We note that our bias measurements  do not show evidence that SMGs at $z$\,$=$\,1.5--2.0  reside in significantly lower mass haloes than the $z$\,$>$\,2 SMGs, which is consistent with the majority of the literature \citep[e.g.][]{chen2016faint,amvrosiadis2019herschel}. In contrast, the \cite{wilkinson2016scuba} sample, that is derived for the same single-dish parent sample as our study, shows a strong evolution in derived halo masses with a  lower absolute bias measurement at $z$\,$<$\,2 than our sample. The strength of this potential `downsizing' behaviour found in \cite{wilkinson2016scuba} was increased by their highest redshift bin at $z$\,$>$\,3 that found SMGs residing in higher mass haloes ($\log_{10}(M_{\rm halo}[{h^{-1}\,\rm M_{\odot}}])$\,$>$\,13). 

We search for evidence of a potential strengthening of the clustering of the higher redshift SMGs by repeating the clustering analysis in our highest redshift bin but extending it out to encompass $z$\,$=$\,2.5--3.5 (increasing the  number of SMGs in the sample to $ N $\,$=$\,218). If SMGs at $z$\,$>$\,3  do reside in significantly more massive haloes, then we would expect to find the estimated bias in the extended redshift bin to increase. However, we estimate a halo mass of $\log_{10}(M_{\rm halo}[{h^{-1}\,\rm M_{\odot}}])$\,$=$\,12.9$^{+0.5}_{-0.6}$, that is just 0.1\,dex above our estimate for the $z$\,$=$\,2.5--3.0 sample (well within 1\,$\sigma$).
Whilst this hints to a potential marginal increase in halo mass for $z$\,$>$\,3 SMGs, we also note that our original estimate is close to the lower limit proposed by \cite{wilkinson2016scuba} given their substantial uncertainties. As the \cite{wilkinson2016scuba} analysis employs probabilistically-identified counterparts for the S2CLS UDS submillimetre sources, in contrast to  the ALMA interferometic identifications used in our analysis, one likely source of this discrepancy comes from the mis-identification of the SMGs. 

We estimate this contamination in the \cite{wilkinson2016scuba} sample by taking their parent SMG catalogue from \cite{chen2016scuba} and applying their same `Class 1' SMG selection, defined as SMGs in regions of the UDS map with both optical and radio coverage (similar to our own selection described above). Of the 645 \cite{chen2016scuba} SMGs used by \cite{wilkinson2016scuba}, just 392 match to an ALMA AS2UDS SMG to within a 1.0 arcseconds matching radius, corresponding to a 42\,\% contamination rate. The contamination rate is highest at the lower redshift end ($z$\,$=$\,1.5--2.0) with $\sim$\,52 per cent of the probabilistically-identified SMGs having no ALMA detection. As the genuine ALMA-detected SMGs are expected to be on average more massive than contaminant mis-identified galaxies, the significantly lower characteristic halo masses predicted by \cite{wilkinson2016scuba} at this redshift range would be a natural consequence of this level of contamination. In the redshift range $z$\,$=$\,2--3, where the \cite{wilkinson2016scuba} clustering measurements best agrees with the literature, the contamination rate is lower at $\sim$\,30 per cent.  A similar  contamination rate of $\sim$\,30 per cent applies at the highest redshift bin of $z$\,$=$\,3--4 where they claim evolution in the halo mass.

In addition to measuring average halo masses for the SMGs at the redshift that they are observed, we can constrain the possible descendants of these SMGs by estimating their present-day halo masses using the median halo growth rates given in \cite{fakhouri2010merger}. As shown in Figure~\ref{fig:Fig4} when applying this median growth rate, the resulting median halo masses for the SMGs in all three redshift bins yield roughly consistent masses of $\log_{10}(M_{\rm halo}[{h^{-1}\,\rm M_{\odot}}])$\,$\sim$\,13.2 at $z$\,$\sim$\,0. The evolved $z$\,$=$\,0 biases and masses for all three SMG redshift bins  are consistent with those expected for the progenitors of local halos that host  2--4\,$L^{\star}$ galaxies, a population that is dominated by passive spheroidal systems \citep{zehavi2011galaxy}.

The high present-day halo masses and the properties of galaxies normally found to populate such haloes, are consistent with a range of other circumstantial evidence linking SMGs with the formation of spheroidal galaxies.  Recent examples include the broad agreement between SMG properties and the scaling relations seen in local spheroids between baryonic surface density and total stellar mass, $\Sigma_{\rm bar}$--$M_\ast$, by \citet{hodge2016kiloparsec} \citep[see also][]{franco2020goods}, and between velocity dispersion and total baryonic mass, $\sigma$--$M_{\rm bar}$, by \cite{birkin2020alma}, as well as their general sizes \citep[e.g.][]{fujimoto2017demonstrating,ikarashi2017very} and environmental trends \citep[e.g.][]{zavala2019gas}. Similarly, \citet{dudzevivciute2020alma} demonstrate that the space density of massive SMGs roughly matches that for the most massive, evolved galaxies (in general agreement with some theoretical simulations, \citealt[][]{mcalpine2019nature}).  Our results on the halo masses of SMGs provide further support for the simple empirical model linking these massive intensely star-forming and metal rich galaxies at high redshift to the early formation and evolution of local, passive spheroids.

For an empirical comparison to other high-redshift galaxy populations we show in Figure~\ref{fig:Fig4} the estimated correlation lengths for a range of previous studies collected from \cite{hickox2012laboca} e.g.\ luminous QSOs \citep{myers2006first,ross2009clustering}, Lyman-break galaxies (LBGs) \citep{adelberger2005possible}, {\it Spitzer} MIPS 24-$\mu$m selected star-forming galaxies \citep{gilli2007spatial}, and optically selected clusters \citep{estrada2009correlation}. We also show the predictions for the correlation length evolution with redshift at varying halo masses using the \cite{peebles1980large} formalism. Our results are again, roughly consistent with \cite{hickox2012laboca}, showing across $z$\,$=$\,1.5--3 that our SMGs have similar correlation lengths to QSOs \citep{myers2009incorporating} whilst, even with the large uncertainties, the AS2UDS SMGs appear to be more strongly clustered than the typical optically selected star-forming populations in the same redshift range \citep[e.g.][]{adelberger2005spatial}. 

\begin{figure}
 \includegraphics[width=\columnwidth]{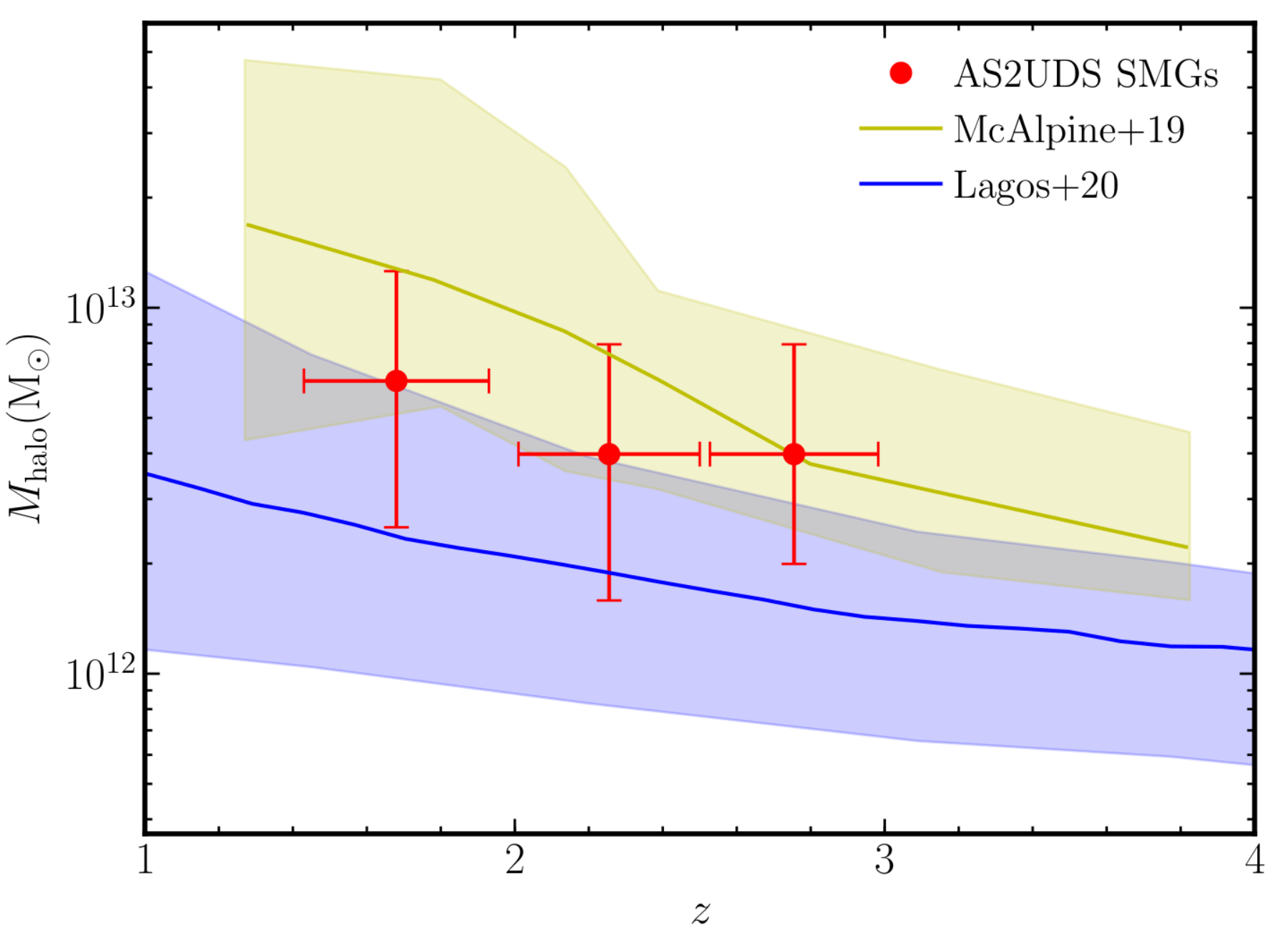}
 \caption{The estimated halo masses for the AS2UDS SMGs compared to two recent theoretical models of SMGs. The two lines show the median values for each model, with the shaded region representing the 16--84\,th percentile range for the semi-analytic model \textsc{shark} \citep{lagos2020physical}, and  the 10--90\,th percentile range in the case of the N-body hydrodynamical simulation \textsc{eagle} \citep{mcalpine2019nature}. Whilst we see general agreement to our observational results, we note that the AS2UDS SMGs are typically brighter $S_{870}$\,$\gtrsim$\,3.6\,mJy SMGs than those predicted by either model.}
 \label{fig:Fig5}
\end{figure}

Turning to theoretical studies, we see that our measured halo masses in the three redshift bins and their lack of evolution is consistent with results from recent simulations of SMGs. Figure~\ref{fig:Fig5} shows the halo masses derived by \cite{mcalpine2019nature} for $S_{\rm 870}$\,$>$\,1\,mJy SMGs from the \textsc{eagle} N-body hydrodynamical simulation. These are well-matched to our results and show only  mild `up-sizing' in halo masses with redshift, consistent with the lack of a strong trend in our data.  We also show the results from \cite{lagos2020physical} using their \textsc{shark} semi-analytic model to reproduce the properties of SMGs. In Figure~\ref{fig:Fig5} we show the halo masses for their  simulated SMGs with $S_{870}$\,$>$\,1\,mJy.  Their model shows a slow increase in halo mass with time, similar to that seen in \textsc{eagle}, but the halo masses they derive are somewhat lower than the observations \citep[and those in][]{mcalpine2019nature}. The lower halo masses in \textsc{shark} could be due to the flux limit, as our AS2UDS sample has an effective flux limit of $S_{870}$\,$\sim$\,4\,mJy, and \cite{lagos2020physical} find that the model SMGs with lower 870-$\mu$m fluxes reside in haloes of lower masses, with $S_{870}$\,$>$\,0.01\,mJy galaxies residing in haloes $\sim$\,0.8 dex less massive than those with $S_{870}$\,$>$\,1\,mJy.  It is possible therefore that the halo masses of a flux-matched \textsc{shark} sample might more closely match our AS2UDS SMGs.

%
%

%
%
\begin{figure*}
 \includegraphics[width=2\columnwidth]{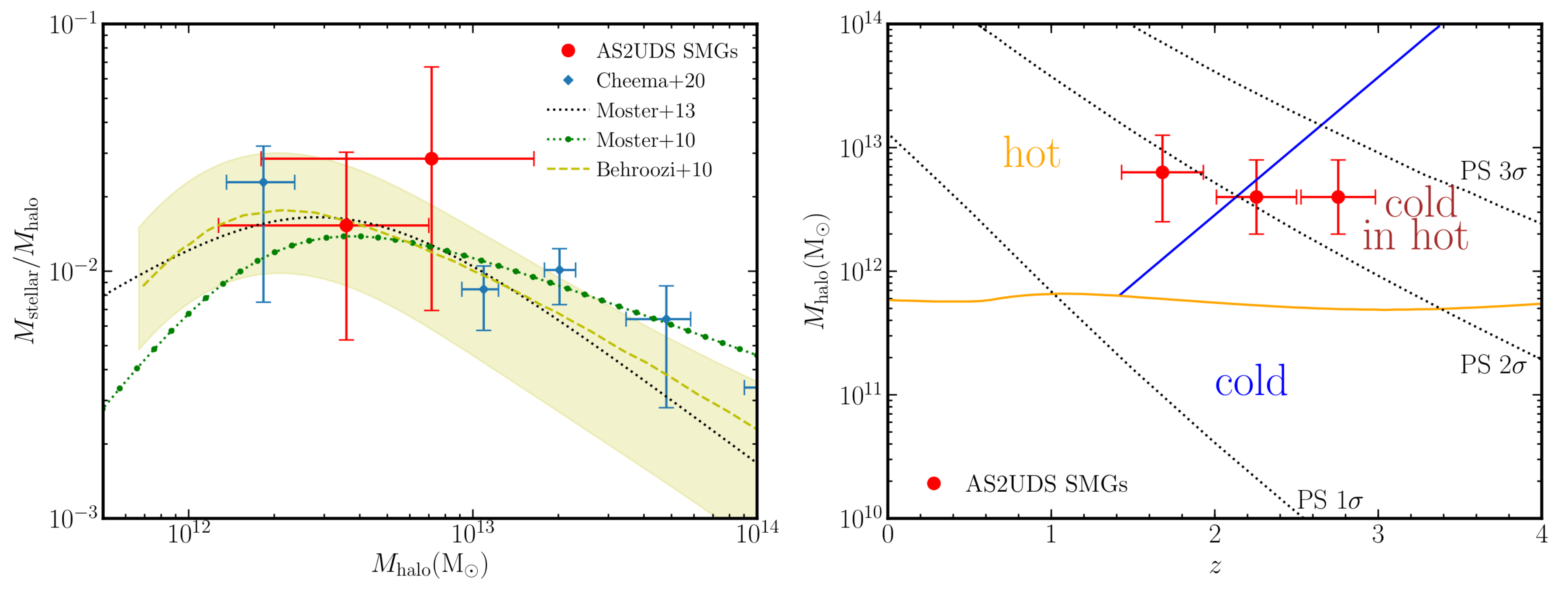}
 \caption{{\it Left:}The stellar mass-halo mass ratio for the AS2UDS SMGs as a function of halo mass. For comparison we show a number of empirical model predictions for central galaxies at $z$\,$=$\,2 from \protect\cite{moster2010constraints,moster2013galactic} with dotted and dot-dashed lines respectively, and a dashed line with the shaded region showing models using the abundance matching from \protect\cite{behroozi2010comprehensive}, where the shaded region shows the 1-$\sigma$ uncertainties. Observational  results for $z$\,$\sim$\,1.6 massive $gzK_s$-selected quiescent galaxies \protect\citep{cheema2020large} are shown by the diamond points. The models show that SMGs, with halo masses $\log_{10}(M_{\rm halo}[{\rm M_{\odot}}])$\,$\sim$\,12.5--12.8 are expected to maximal stellar mass-halo mass ratios and thus represent the peak efficiency of star formation at their redshifts.
 {\it Right:} The redshift-binned SMG characteristic halo masses in comparison to the \protect\cite{dekel2006galaxy} schematic for the thermal properties of gas flowing  onto galaxies. Below the almost horizontal orange line the galaxy discs are fed by cold streams conducive to future star formation. The diagonal solid blue line is the upper limit for a galaxy's mass at the critical redshifts where the cold streams can still penetrate the hot shock heated halos and thus star formation can still occur.  Our estimated halo masses for the SMGs lie around this boundary, suggesting that they may represent the most massive, common haloes onto which gas can cool to fuel star formation.
 The dotted lines are from the Press--Schechter estimates for halo formation masses and show the estimated percentage of the total halo mass at a given redshift that resides in haloes of mass greater than $M_{\rm halo}$ (i.e.\ 1\,$\sigma$\,$=$\,22 per cent, 2\,$\sigma$\,$=$\,4.7 per cent and 3\,$\sigma$\,$=$\,0.3 per cent). }
 \label{fig:Fig6}
\end{figure*}

\subsection{Stellar-to-Halo mass ratio}\label{sec:shmr}

Finally, we investigate the stellar-to-halo mass ratio (SHMR) for the AS2UDS SMGs as a function of their halo masses. SHMR is a measure of the efficiency with which these galaxies  form stars. To estimate the SHMR  we take the stellar mass estimates for the SMGs from \cite{dudzevivciute2020alma} that are estimated using SED fitting with \textsc{magphys} \citep{da2008simple,da2015alma,battisti2019magphys+}.  

To estimate the SHMR we first separate our SMGs into two bins of stellar mass (to minimise variations in the halo mass estimates), split at the median stellar mass of the sample, $\log_{10}(M_{\star}[{\rm M_{\odot}}])$\,$\sim$\,11.1.  The  mass  for each bin is the median stellar mass with an uncertainty given by  the 10--90\,th percentiles of the distribution for each bin.  For both bins the characteristic halo masses are derived in the same manner as above and the resulting SHMR ratios are shown in Figure~\ref{fig:Fig6}.  We estimate $\log_{10}(M_{\rm stellar}/M_{\rm halo})$\,$=$\,$-$1.8$^{+0.4}_{-0.6}$ for our lower mass bin and $\log_{10}(M_{\rm stellar}/M_{\rm halo})$\,$=$\,$-$1.5$^{+0.4}_{-0.6}$ for the higher mass, both estimates being consistent within the significant uncertainties.

For comparison, we also show observational estimates  of the SHMR for $gzK$-selected quiescent galaxies at $z$\,$\sim$\,1.6 from \cite{cheema2020large}, that as we mentioned above, could be immediate descendants of our SMG population. We see that the SHMR for the SMGs are broadly consistent with the maximum SHMR inferred for these quiescent galaxies, supporting the presence of a peak in the SHMR in haloes of mass $\log_{10}(M_{\rm halo}[{\rm M_{\odot}}])$\,$\sim$\,12.5, characteristic of SMGs. 

We also show in 
Figure~\ref{fig:Fig6} the predicted SHMR ratio tracks from theoretical models \citep{behroozi2010comprehensive,moster2010constraints,moster2013galactic}. 
These models show a broad peak in the SHMR arising from the influence of
two competing feedback processes in high- and low-mass haloes.  In the higher  mass haloes the 
theoretical models suggest star formation is suppressed due to AGN heating of the halo gas content that prevents cooling, removing the reservoir of cold gas needed to fuel star formation  \citep[e.g.][]{bower2006breaking,croton2006many}. While in  lower mass halos the shallower gravitational potential well is thought to be insufficient to contain gas being ejected from supernovae winds \citep{larson1974effects,dekel1986origin,puchwein2013shaping}.
Compared to these models, our SHMR ratios are in agreement with the theoretical predictions, with both stellar mass bins resulting in a SHMR at the peak of the model predictions. 
 As noted, these models show suppression of the star formation for lower- and higher-mass haloes with a peak at an intermediate mass of $\log_{10}(M_{\rm halo}/{M_{\odot}})$\,$\sim$\,12.5, i.e.\ the halo mass range where the predicted physical processes that suppress star formation are least efficient, matching the behaviour we see in the observations.
 
 The association of SMGs with the era of peak efficiency in the formation of massive galaxies appears to be the fundamental basis for much of their behaviour, including
 their star-formation rates, stellar and halo masses and redshift distribution \citep{dudzevivciute2020alma}.  Theoretical studies, such as \citet{white1991galaxy,dekel2006galaxy}, have related the star-formation activity in galaxies to the masses of their host haloes and their ability to accrete cold gas using simple recipes.  \citet{dekel2006galaxy} identify different regimes for gas cooling in haloes as a function of redshift and halo mass:  in haloes with masses below $\log_{10}(M_{\rm halo}/{M_{\odot}})$\,$\sim$\,12, gas can cool from the intragalactic medium onto the central galaxy at all redshifts.  However, for more massive haloes a shock forms in the halo that increasingly limits the ability of streams of cold gas to be accreted onto the central galaxy at lower redshifts. This behaviour is illustrated in Figure~\ref{fig:Fig6}, that shows the boundaries of the various regimes as well as the collapse redshifts for haloes of different masses as indicated by the rarity of the fluctuations they represent based on the Press-Schecter formalism (which gives some indication of the likely rarity of haloes with a given mass as a function of redshift). 
 In \citet{dekel2006galaxy} the disruption of the cold streams in massive haloes occurs at a mass scale that is a multiple of the characteristic halo mass at that epoch, reflecting the influence of the local environment and growing large-scale structures on the halo accretion.  This results in the diagonal boundary line shown in Figure~\ref{fig:Fig6} -- between regimes at high and low redshifts where the cold streams can or cannot  feed the central galaxy. 
 
  We indicate on Figure~\ref{fig:Fig6} the halo masses and median redshifts for SMGs in the three redshift ranges analysed in \S~\ref{sec:evolution}.  These measurements are broadly consistent with these SMGs lying near the boundary defining  the most massive galaxies where the  cold streams can still feed the star-formation activity in galaxies.    These galaxies thus represent the most massive  galaxies that can continue to support significant star-formation rates fueled by the accretion of gas supplies from the surrounding intragalactic medium.   This model also naturally explains the peak in the redshift distribution of dust-mass-selected samples of SMGs at $z$\,$\sim$\,2--3, with an exponential decline at higher redshifts, $z$\,$\geq$\,3--4 \citep{dudzevivciute2020alma}, that corresponds to the increasing rarity   of such massive halos at higher redshifts.

%
%
%

\section{Conclusions} \label{sec:conclusions}

We have measured the clustering strength of the largest sample of interferometrically identified SMGs in a single contiguous field. We use Monte Carlo methods to incorporate the complete photometric redshift PDFs for both the main SMG samples and the $K$-selected field  galaxy sample, into the calculation of the projected cross-correlation functions. The main results of our clustering analysis are as follows:
\begin{itemize}
  
  \item Across the entire redshift range considered ($z$\,$=$\,1.5--3.0) we find an SMG correlation length of $r_0$\,$=$\,6.6$^{+1.9}_{-1.9}$\,$h^{-1}$\,Mpc. This is consistent with previous studies of smaller samples of single-dish detected SMGs that show SMGs to be more strongly clustered than typical star-forming galaxies at their redshift and more similar to the clustering strength seen for luminous QSOs (or `bright quasars'). From linearly scaling the dark matter projected correlation function we derive dark matter halo masses for $z$\,$=$\,1.5--3 SMGs of $\log_{10}(M_{\rm halo}[{h^{-1}\,\rm M_{\odot}}])$\,$=$\,12.6$^{+0.3}_{-0.4}$. 
  
  \item We split our sample into three redshift bins $z=$\,1.5--2.0, 2.0--2.5, and 2.5--3.0 and find, contrary to some previous studies, no significant evolution in the dark matter halo masses with redshift. The SMGs in each redshift bin reside in  haloes with median masses of $\log_{10}(M_{\rm halo}[{h^{-1}\,\rm M_{\odot}}])$\,$\sim$\,12.7 and from the \cite{fakhouri2010merger} median halo growth rates we estimate that the typical $z$\,$=$\,1.5--3.0 SMG will reside in haloes with mass $\log_{10}(M_{\rm halo}[{h^{-1}\,\rm M_{\odot}}])$\,$\sim$\,13.2 by the present day that is consistent with the picture of SMGs evolving into local massive passive elliptical galaxies \citep{sanders1988ultraluminous}. 
  
  \item Exploiting the stellar mass estimates for the SMGs from \cite{dudzevivciute2020alma} we split the AS2UDS sample into two stellar mass bins and calculated their respective characteristic halo masses. The stellar to halo mass ratio for these sub-samples are consistent with the theoretical models, with the SMGs lying at the peak of the stellar-to-halo mass ratio for the models.  This  suggests that SMGs are amongst the most efficient galaxy populations in terms of the conversion of baryons into stellar mass.

  \item We compare the estimates of the halo masses for SMGs as a function of redshift to a simple model that describes the bimodality in the local galaxy population through a dichotomy in the mode of gas accretion, driven by the presence of a stable shock in gas accreting in more massive halos.  We show that the SMGs fall near the boundary where cold gas streams can still be accreted onto the central galaxies in the most massive haloes.  This would naturally explain several characteristics of the SMG population, including their intense star-formation rates, masses and redshift distribution, as they  represent the most massive  galaxies that can still support their star-formation activity through accretion of gas from the intragalactic medium.

\end{itemize}

\section*{Acknowledgements}

All Durham co-authors acknowledge financial support from  STFC  (ST/T000244/1). AA is supported by ERC Advanced Investigator grant, DMIDAS [GA 786910], to C.S.\ Frenk. CCC acknowledges support from the Ministry of Science and Technology of Taiwan (MOST 109-2112-M-001-016-MY3). KEKC acknowledge support from the UK Science and Technology Facilities Council (STFC) (grant number ST/R000905/1) and a Royal Society Leverhulme Trust Senior Research Fellowship (grant number RSLT SRF/R1/191013). JLW acknowledges support from an STFC Ernest Rutherford Fellowship (ST/P004784/1 and ST/P004784/2).



\section{Data Availability}

The data underlying this article are available in the JCMT, ALMA and ESO archives.

\bibliographystyle{mnras}
\bibliography{AS2UDS_Clustering} 



\appendix

\bsp	
\label{lastpage}
\end{document}